\documentclass[conference]{IEEEtran}
\IEEEoverridecommandlockouts
\usepackage{cite}
\usepackage{amsmath,amssymb,amsfonts}
\usepackage{algorithmicx}
\usepackage{algorithm}
\usepackage{graphicx}
\usepackage[lofdepth,lotdepth]{subfig}
\usepackage[font=footnotesize,labelfont=bf]{caption}
\usepackage{textcomp}
\usepackage{ctable}
\usepackage{tabularx}
\usepackage{xcolor}
\usepackage{float}
\usepackage{enumitem}
\usepackage{booktabs}
\usepackage{multirow}
\usepackage[square,numbers]{natbib}
\usepackage{cleveref}

\usepackage{amsmath}
\usepackage{algorithm}
\usepackage[noend]{algpseudocode}

\makeatletter
\def\BState{\State\hskip-\ALG@thistlm}
\makeatother

\def\BibTeX{{\rm B\kern-.05em{\sc i\kern-.025em b}\kern-.08em
    T\kern-.1667em\lower.7ex\hbox{E}\kern-.125emX}}

\DeclareMathOperator*{\argmin}{arg\,min}

\usepackage{fancyhdr,lipsum}
\fancypagestyle{firstpage}{
  \fancyhf{}
  \fancyhead[C]{To appear at the 2020 International Joint Conference on Neural Networks (IJCNN), Glasgow, Scotland, July 2020.}
  \fancyfoot[C]{\thepage}
}

\begin{document}
\title{NeuroAttack: Undermining Spiking Neural Networks Security through Externally Triggered Bit-Flips \vspace*{-11pt}}

\author{\IEEEauthorblockN{Valerio Venceslai$^{1,2,*}$\thanks{*These authors contributed equally to this work.}, Alberto Marchisio$^{1,*}$, Ihsen Alouani$^3$, Maurizio Martina$^2$, Muhammad Shafique$^1$}
\IEEEauthorblockA{\textit{$^1$Technische Universität Wien, Vienna, Austria}\\
\textit{$^2$Politecnico di Torino, Turin, Italy}\\
\textit{$^3$Université Polytechnique Hauts-De-France, Valenciennes, France}\\
\textit{Email: s254591@studenti.polito.it, \{alberto.marchisio, muhammad.shafique\}@tuwien.ac.at,}\\
\textit{ihsen.alouani@uphf.fr, maurizio.martina@polito.it
}}
\vspace*{-25pt}}

\maketitle
\thispagestyle{firstpage}

\begin{abstract}
Due to their proven efficiency, machine-learning systems are deployed in a wide range of complex real-life problems.  
More specifically, Spiking Neural Networks (SNNs) emerged as a promising solution to the accuracy, resource-utilization, and energy-efficiency challenges in machine-learning systems. While these systems are going mainstream, they have inherent security and reliability issues.
In this paper, we propose NeuroAttack, a cross-layer attack that threatens the SNNs integrity by exploiting low-level reliability issues through a high-level attack. Particularly, we trigger a fault-injection based sneaky hardware backdoor through a carefully crafted adversarial input noise. Our results on Deep Neural Networks (DNNs) and SNNs show a serious integrity threat to state-of-the art machine-learning techniques.

\end{abstract}


\section{Introduction}



Deep Neural Networks (DNNs) are known to be resilient to numerical perturbations and architectural imprecision~\cite{8342139}\cite{Marchisio2020ReD-CaNe}\cite{Shafique2020RobustML}\cite{Zhang2019RobustML}. This is demonstrated through an established performance even after aggressive pruning~\cite{Marchisio2018PruNet}, quantization~\cite{Marchisio2020Q-CapsNets}, and other compression techniques~\cite{Han2016DeepCompression}\cite{Hanif2018X-DNNs}, which significantly reduce the number of parameters in the network. However, recent works~\cite{Hanif2019SalvageDNN}\cite{Hoang2020FTClipAct}\cite{iccd18}\cite{dant} have shown that these networks are vulnerable to surgical bit-flips in specific locations. Moreover, system-level threats called adversarial attacks~\cite{Goodfellow} have shown effective ability to induce behavioral anomalies in DNNs. In fact, DNNs are vulnerable to malicious inputs modified to yield erroneous labels, while being undetectable to human observers~\cite{Hanif2018RobustML}\cite{Marchisio2019DL4EC}. In safety-critical applications such as transportation systems, adversarial examples could be a non-negligible threat to public safety.
For this reason, attacks and defenses on adversarial examples have drawn great attention in the scientific community. 
On the other hand, due to the ubiquity of machine-learning, attacks from the supply chain such as hardware Trojans emerged as a threat to DNNs security. In~\cite{FICNN}, the authors use fault-injection techniques on SRAM or DRAM to alter the single bit value or few bit values in memory thereby leading to misclassification.

Spiking Neural Networks (SNNs) provide a biologically plausible alternative to DNNs, because the neuron model as well as the event-based communication model between neurons resemble to the current understanding of the human brain's functioning. Compared to DNNs, SNNs show a different response to the adversarial attacks~\cite{Marchisio2019SNNAttack}. Moreover, due to their asynchronous and spike-based propagation, the SNNs are naturally more energy-efficient than DNNs when deployed in the hardware, as shown by neuromorphic chips like Intel Loihi~\cite{Davies2018Loihi} and IBM TrueNorth~\cite{Merolla2014TrueNorth}.


Towards this, the focus of our paper is to show a new attack vector that threatens the integrity of both the DNNs and SNNs. We propose a cross-layer attack against neural networks that transforms a circuit-level vulnerability to a system-level security flaw. We exploit memory bit-flips in neural networks synapses' weights through a hardware Trojan triggered using a surgical adversarial attack.

\textit{To the best of our knowledge, this is first end-to-end attack against SNNs that exploits circuit-level backdoor through a high-level input pattern.}

In summary, the contributions of our paper are as follows:

\begin{itemize}
    \item We analyze the resilience of SNNs to errors. 
    \item We propose a methodology for triggering a bit-flip attack remotely through an adversarial input pattern. 
    \item We introduce \textbf{\emph{NeuroAttack}}, a hardware Trojan triggered by an input noise. We design and compare different versions of the noise pattern that triggers the Trojan.  
    \item We show the practicality of NeuroAttack on DNNs and SNNs, by converting pre-trained DNNs into the spike domain. 
\end{itemize}


\section{Background and Related Work}
\label{sec:background}

\subsection{Spiking Neural Networks}

Spiking Neural Networks (SNNs) are considered as the 3rd generation neural networks. The previous generations employed continuous values for the output signals of the neurons, whereas SNNs use spike trains to encode the information. Therefore SNNs, for their binary (spiking or no spiking) operation, lend themselves well to fast and energy-efficient implementation on hardware devices~\cite{Hazan_2018}. Each incoming signal from an input neuron, which is encoded in the SNN technology as a spike train, is multiplied by the weight of the synapses, and all the results are added together to produce the so-called \textit{membrane potential} $V_{m}$, expressed as:
\[
V_{m} = \sum_{i=1}^N w_{i} \cdot s_{i},
\]
where \textit{N} is the number of input synapses. When the membrane potential reaches a particular value, called \textit{threshold}, the output neuron ``spikes", or ``fires".

There are different ways in which the continuous values can be coded as spikes in time domain. The most commonly used are \textit{rate coding} and \textit{time coding}. In the first case, the information is encoded by the number of spikes per second, i.e., an higher number of spikes per second refers to an higher analog value. In this case, the spike rate is determined by the mean rate of a Poisson process~\cite{DBLP:journals/corr/abs-1903-12272}. Moreover, the pixels of the images are converted to a constant current entering in the input neurons, so that they will spike at constant rates depending on the input pixel intensity. The \textit{time coding} can be implemented in different ways, for example the \textit{latency coding}, in which the analog value is inversely proportional to the spiking delay of the neuron.

\begin{figure}[ht!]
\centering
\vspace*{-13pt}
\includegraphics[width=\linewidth]{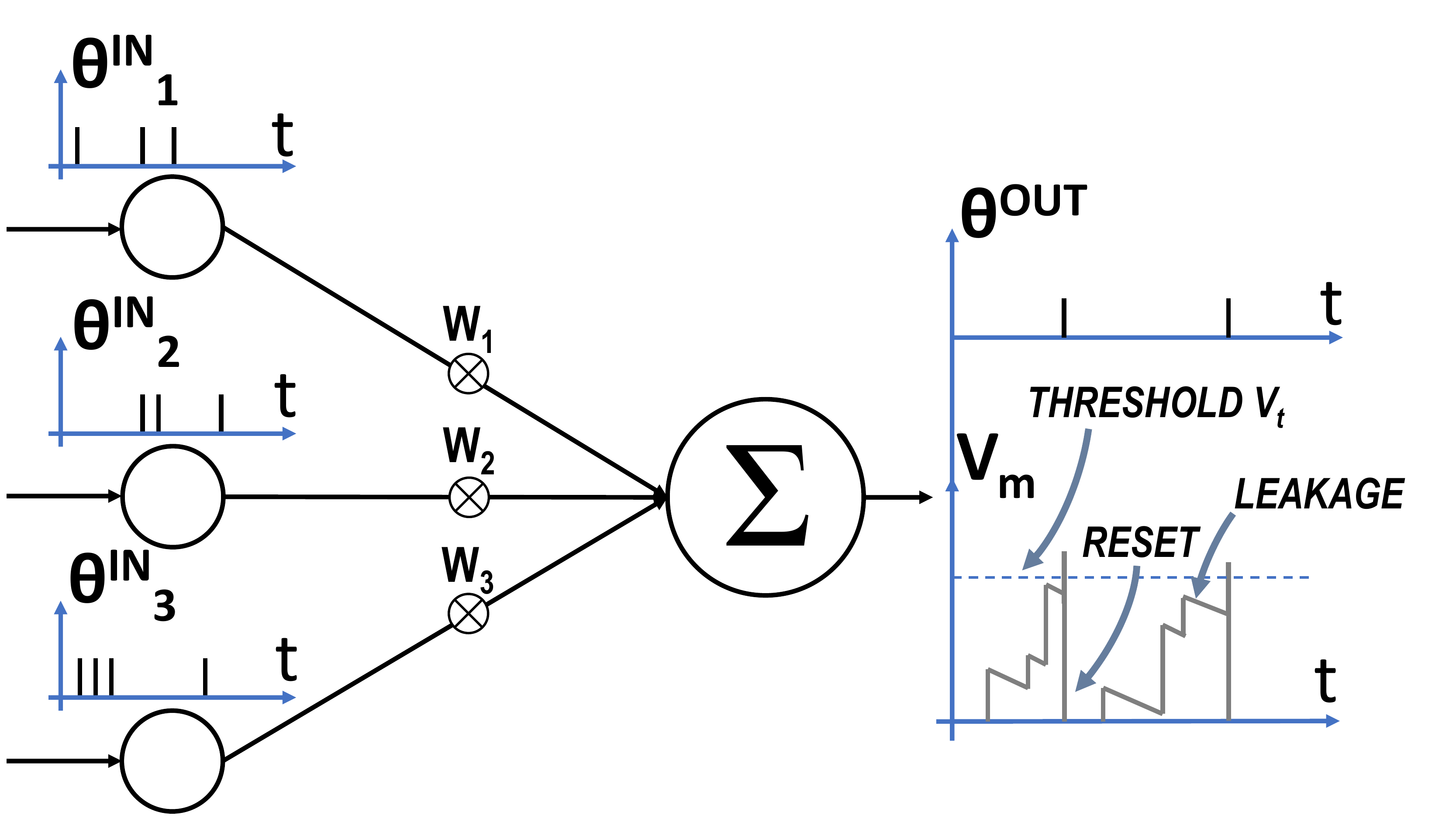}
\caption{Input and output spikes, referred to the membrane potential for a simple LIF model.}
\label{fig:LIF}
\end{figure}

Many different models for the spiking neurons have been studied. These models must be at the same time (1) biologically accurate and capable of producing rich patterns, and (2) computationally simple. The Hodgkin-Huxley biologically-accurate model~\cite{Hodgkin-Huxley} is computationally expensive, whereas, on the other hand the \textit{Leaky Integrate and Fire} (LIF) model~\cite{LIF_neuron_model} gives the opportunity, for its simplicity, to process lots of neurons in real-time but its biological plausibility is very low compared to the Hodgkin-Huxley's model. Other models have been developed to make a compromise between the two extremes. An example of such a tradeoff is the Izhikevich model~\cite{1257420}. However, we take advantage of the simple LIF model (shown in Figure~\ref{fig:LIF}) to explain in the details the working principles of a SNN, as it has been deployed in real-world neuromorphic processors. 

When a spike inputs the neuron, the associated synaptic weight $w_{i}$ will be integrated on the membrane. When the membrane potential $V_{m}$ overcomes a threshold $V_{t}$, the neuron fires and resets its membrane potential to a value $V_{R}$, which is considered to be zero in Figure~\ref{fig:LIF}. In addition, due to leakage, the membrane potential decreases continuously at the leak rate between two input spikes~\cite{Bouvier}. The sub-threshold dynamics of LIF spiking neuron can be formulated as follows: 
\[
\tau_{m} \dfrac{dV_{m}}{dt}=-V_{m}+I(t),
\]
where $V_{m}$ is the membrane potential and $\tau_{m}$ is the time constant for the membrane potential leakage~\cite{DBLP:journals/corr/abs-1903-06379}.
Local learning rules for unsupervised learning can be used to train the network, as can be done also in the recent Loihi neuromorphic processor~\cite{Davies2018Loihi}. The \textit{Spiking Time Dependent Plasticity} (STDP) local learning rule can be applied. The goal of such a rule is to strengthen the synaptic weight of two neurons whose spiking activity happens in a highly-correlated causal dependency order, and to weaken it otherwise~\cite{STDP}. However, learning through the unsupervised learning rules is found to be effective just for shallow networks~\cite{Lee2018CNNSTDP}. On the contrary, the backpropagation mechanism used to train DNNs cannot be applied as-is, due to the non-differentiabile nature of the spiking function ~\cite{DBLP:journals/corr/abs-1903-06379}. To overcome this problem two solutions are typically employed: (1) take advantage of an approximate derivative method, or (2) convert offline trained DNNs to SNNs. The first solution has been extensively studied in many works~\cite{BOHTE200217}\cite{DBLP:journals/corr/abs-1903-06379}\cite{DBLP:journals/corr/LeeDP16}\cite{SLAYER}. The second solution is exploited in the following discussions. The neural networks are described as Keras models, trained as DNN and then converted to SNN by means of the SNNtoolbox~\cite{10.3389/fnins.2017.00682}, and implemented by means of spiking neuron's simulators through \textit{rate encoding}. A built-in simulator based on Keras, i.e., INIsim, is used, which features the simple LIF neuron model. The duration of the simulation is set to 50 milliseconds, one millisecond for each time step while the other parameters are left with the default values.

\subsection{Adversarial Attacks}

An adversary, using information learnt about the structure of the classifier, tries to craft the perturbations added to the input to cause its misclassification, i.e., its incorrect classification. For explanation purposes, we consider a generic DNN for image classification. Given an original input image $x$ and a target classification model $ C(.) $, the problem of generating an adversarial example $x^{adv}$ can be formulated as a constrained optimization~\cite{pbform}:

\vspace*{-10pt}
\begin{align*}
x^{adv} =  \argmin_{x^{adv}} \mathcal{D}(x,x^{adv}), 
   s.t.~ \\
   C(x) = l,                                 
           C(x^{adv}) = l^{adv},                     
           l \neq l^{adv}                         
\end{align*}

Where $\mathcal{D}$ is a distance metric used to quantify the similarity between two images, and the goal of the optimization is to minimize the added noise, typically to avoid the detection of the adversarial perturbations. $l$ and $l^{adv}$ are the two labels of $x$ and $x^{adv}$, respectively. Here, $x^{adv}$ is considered as an adversarial example if and only if the label of the two images are different ($ C(x) \neq  C(x^{adv}) $) and the added noise is bounded ($\mathcal{D}(x,x^{adv}) < \epsilon $ where $\epsilon \geqslant 0 $).

\subsection{Fault-Injection}

The outputs of a DNN depend on both the input images and its internal parameters. By inserting errors in the internal parameters of a network, it is possible to misclassify a given input image. Since the parameters of the network, when implemented in hardware, are stored in memory units as SRAM or DRAM, with the development of precise memory fault-injection techniques, such as laser beam fault-injection~\cite{laserfaultinjection} and row hammer attack~\cite{rowhammer}, it is possible to launch effective fault-injection attacks on DNNs~\cite{FICNN}. Shattering the accuracy of a DNN in a significant way, with a low amount of faults, is a challenging task. This is due to the high resilience of neural networks which will be analyzed in section~\ref{sec:bit_flip}. Towards this, an efficient fault-injection technique will be used in Section~\ref{subsec:grad_desc}, and it will be shown that few tens of faults (bit-flips), associated to network's internal parameters, are sufficient to cause a considerable reduction of performances. The results of this analysis will be used to build up an efficient attack methodology through the hypothesis of an hardware Trojan insertion in the supply chain plus a well-crafted input Trojan trigger pattern, which can threaten the security properties of both the DNNs and the SNNs.
Unlike previous works, our NeuroAttack is a \textbf{cross-layer} attack that exploits a hardware backdoor through a carefully crafted adversarial input noise. 
\section{Bit-flip Resilience Analysis of SNNs}
\label{sec:bit_flip}

\subsection{Statistical Analysis of Random Bit-Flip}
In this section, we analyze the resilience of SNNs to random bit-flips in its internal parameters. Two different networks, whose structures are reported in Table~\ref{tab:MLP} and Table~\ref{tab:LeNet}, have been chosen. 

\begin{table}[H]
	\caption{Structure of the Multilayer Perceptron network.}	
	\centering
	\begin{tabularx}{3.1cm}{cc}
		\specialrule{.2em}{.1em}{.1em}
		\textbf{Layer} & \textbf{Output shape}\\
		\specialrule{.2em}{.1em}{.1em}
		Input  & 784\\
		Dense  & 1200\\
		Dense  & 1200\\
		Dense  & 10\\
		\specialrule{.2em}{.1em}{.1em}
	\end{tabularx}
	\label{tab:MLP}
	\vspace*{-10pt}
\end{table}

\begin{table}[h!]
	\caption{Structure of the LeNet network~\cite{LeCun1998LeNet}.}	
	\centering
	\begin{tabularx}{9cm}{cccccc}
		\specialrule{.2em}{.1em}{.1em}
		\textbf{Layer} & \textbf{Output shape} & \textbf{Output maps} & \textbf{Kernel size} & \textbf{Strides}\\
		\specialrule{.2em}{.1em}{.1em}
		Input  & (28, 28, 1) & - & - & - \\
		Conv2D  & (28, 28, 32) & 32 & (5,5) & (1,1)\\
		MaxPool2D  & (14, 14, 32) & - & - & (2,2)\\
		Conv2D  & (10, 10, 48) & 48 & (5,5) & (1,1)\\
		MaxPool2D  & (5, 5, 48) & - & - & (2,2)\\
		Dense  & 256 & - & - & -\\
		Dense  & 84 & - & - & -\\
		Dense  & 10 & - & - & -\\
		\specialrule{.2em}{.1em}{.1em}
	\end{tabularx}
	\label{tab:LeNet}
	\vspace*{-5pt}
\end{table}

The first one is the so called \textit{Multilayer Perceptron} (MLP). The perceptron is a basic neuron, which receives as input the signals multiplied by the synaptic weights. These signals are summed together with a bias $\theta$, and a non-linear function is applied~\cite{266645}, as expressed by the following formula:
\[
f\left(\sum_{i=1}^N x_{i} \cdot w_{i} + \theta\right).
\]
These neurons are connected in a dense (or fully-connected) fashion, so that each neuron in layer \textit{l} receives as inputs the outputs of each neuron in the previous layer \textit{l-1}. The amount of synapses and related weights connecting one layer to the previous one is given by $n_{l-1} \cdot n_{l}$, where $n_{l}$ is the amount of neurons in a given layer \textit{l}. For instance, for a simple 4 layer MLP, like the one in Table~\ref{tab:MLP}, the number of parameters is about 2 millions. This huge amount of parameters is related to an inherent resilience of DNNs to errors or approximations, as it has been studied in prior works~\cite{8342139}\cite{7551399}\cite{8465834}. 

With \textit{Convolutional Neural Networks} (CNNs), additional types of layers are introduced, i.e., the \textit{convolutional layers} to extract features from the input image and the \textit{pooling layers} to reduce the size of the data. The so-called \textit{feature maps} of the convolutional layers sweep the input image with a certain stride, and have shown excellent capabilities to extract features in the images given as inputs. This trait led to reach an outstanding performance in many image-recognition and classification tasks. One example of CNN is the LeNet-5, whose structure is shown in Table~\ref{tab:LeNet}. It achieves excellent capabilities in classifing images belonging to the MNIST dataset.

\begin{figure}[t]
\centering
\includegraphics[width=90mm]{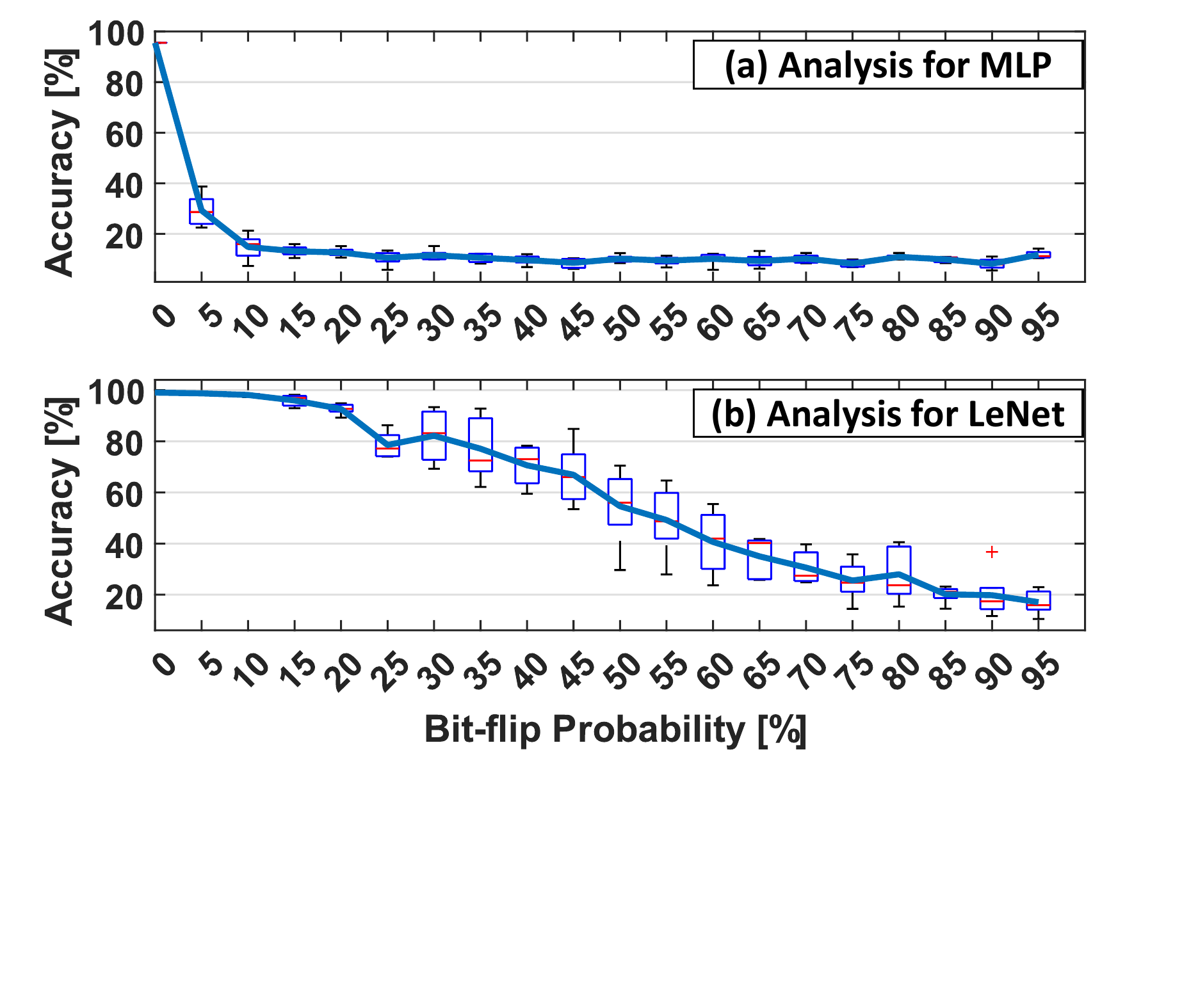}
\vspace*{-68pt}
\caption{Accuracy vs bit-flip probability for (a) MLP, and (b) LeNet network.}
\label{fig:boxplot}
\vspace*{-15pt}
\end{figure}

The two networks have been trained for 30 epochs to reach the top accuracy of 95.54\% and 99.05\% on the MNIST dataset for the MLP and the LeNet, respectively. Weights and biases are then quantized to 8 bits. The first investigation is a statistical analysis of both networks. The \textit{bit-flip probability} is set between 0\% and 95\% to have 20 different points, and it represents the probability for which a weight is subjected to bit-flip. The results are averaged over 5 different iterations. 
The results of accuracy against the \textit{bit-flip probability} for both the MLP and the LeNet are shown in Figures~\ref{fig:boxplot}-a and~\ref{fig:boxplot}-b, respectively.



\begin{figure}[ht!]
\centering
\includegraphics[width=\linewidth]{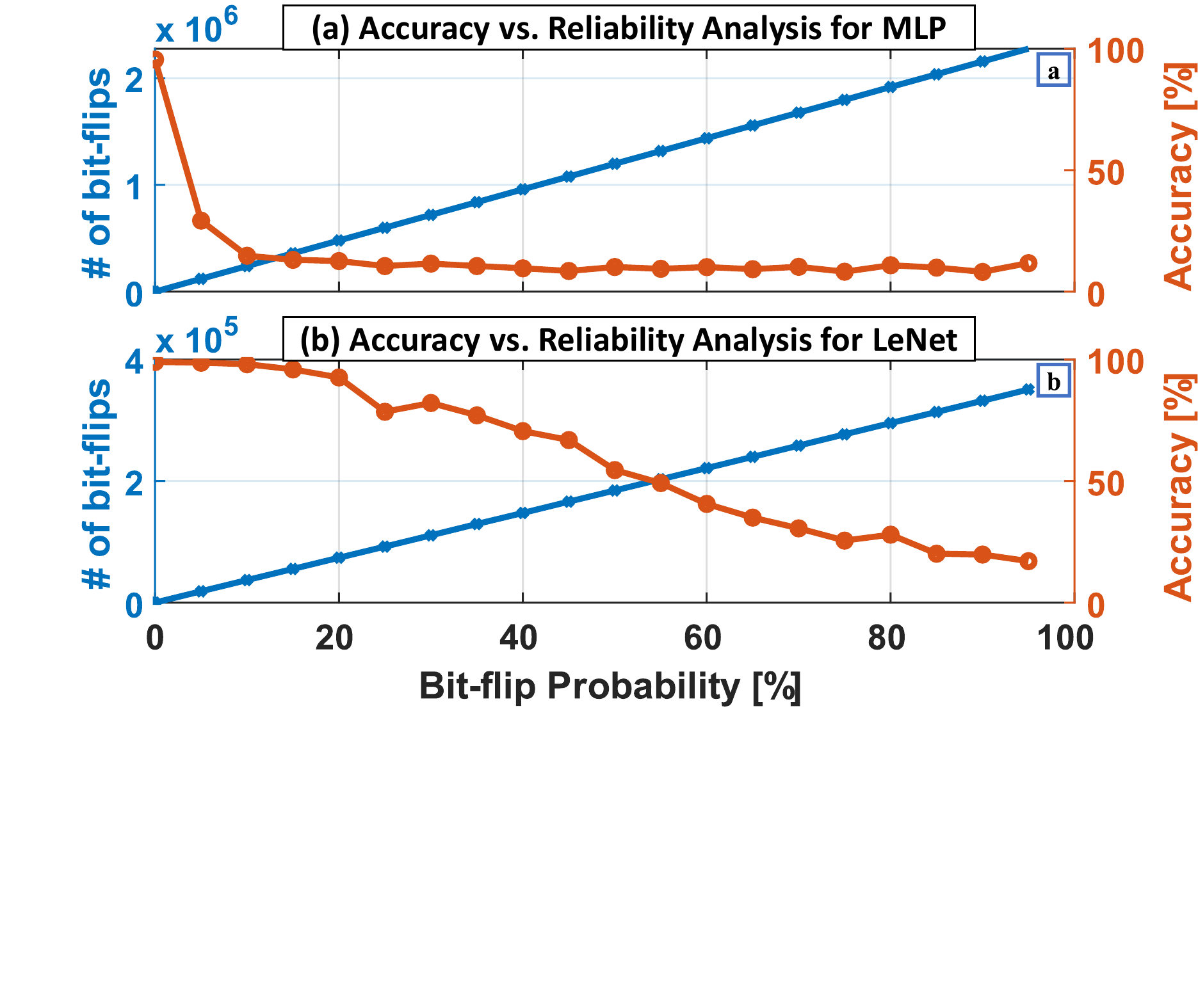}
\vspace*{-70pt}
\caption{Accuracy and number of bit-flips vs bit-flip probability for (a) MLP and (b) LeNet network.}
\label{fig:merge2}
\vspace*{-25pt}
\end{figure}

These results show that in the MLP, the accuracy is reduced significantly also for a low \textit{bit-flip probability}. However, for networks with huge amount of parameters, a higher number of parameters undergo bit-flip also for low values of \textit{bit-flip probability}. The situation is clear looking at Figure~\ref{fig:merge2}-a and Figure~\ref{fig:merge2}-b which depict the average accuracy (red line, right axis) compared to the average number of bits flipped (blue line, left axis), for MLP and LeNet respectively. The number of bits flipped with the same bit-flip probability appear to be at least one order of magnitude less in the LeNet with respect to the MLP. This analysis shows the high resilience of a neural network whose performance is degraded just for a huge amount of errors in the network parameters. However, these networks, as demonstrated in the following section, are resilient only for probabilistic attacks, while showing very different behavior in case of well-targeted errors that can be applied by an adversary.



\subsection{Bit-Flip with Gradient Search Algorithm}\label{subsec:grad_desc}



\begin{figure}[ht!]
\centering
\vspace*{-135pt}
\includegraphics[width=\linewidth]{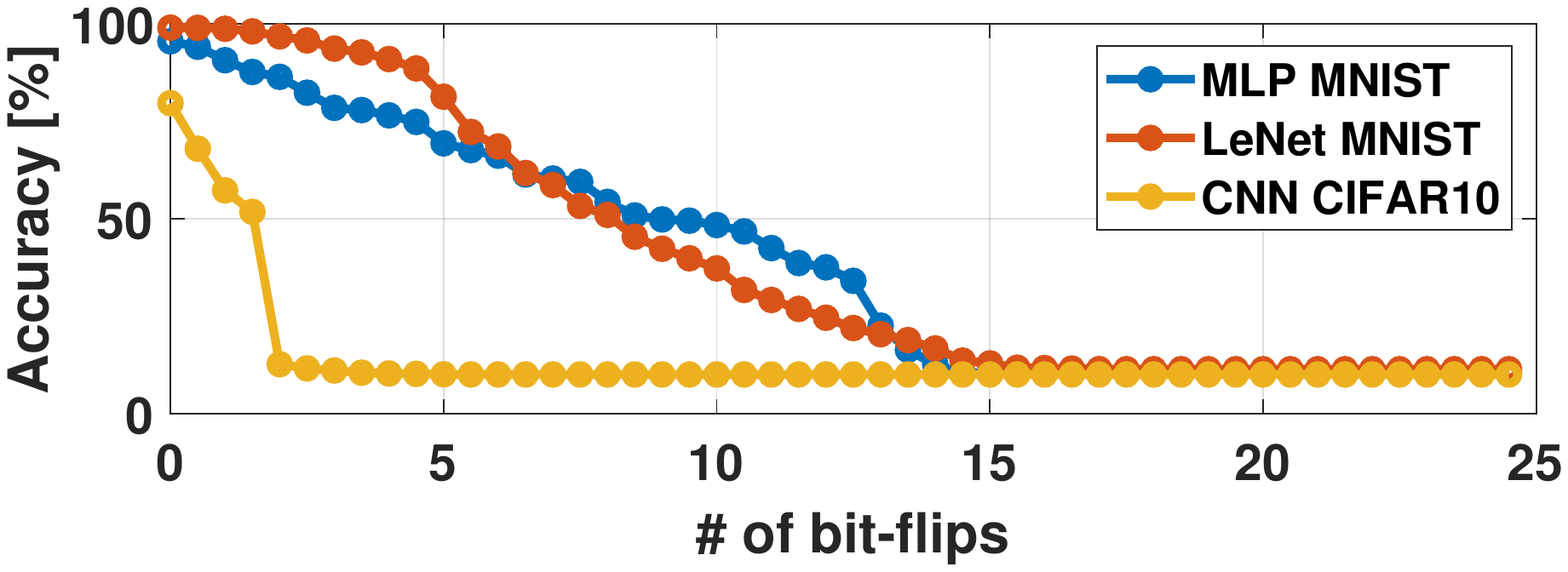}
\vspace*{-150pt}
\caption{Accuracy vs number of bit-flips for MLP@MNIST, LeNet@MNIST and CNN@CIFAR10.}
\label{fig:CIFAR10_MLP_LeNet}
\vspace*{-15pt}
\end{figure}

\textbf{Analysis for the MINIST Dataset}: In this section, we describe a way to reduce the accuracy of a network by applying errors on the lowest possible amount of bits. The gradients of the loss function with respect to the parameters of the network are analyzed in a similar way to what is done during the learning, while taking an inspiration from the work of~\cite{DBLP:journals/corr/abs-1903-12269}. The computation of gradients returns a list of n-dimensional arrays of the same shape of the parameters. The highest gradient in absolute value is taken and the corresponding parameter is considered as the target parameter. One of the bits of the target parameter is flipped to have the maximum reduction of accuracy. The target parameter is then masked, so that it is not considered at the next iteration.
The results show that the accuracy is highly reduced for very low number of bit-flips for the MLP (see the blue line in Figure~\ref{fig:CIFAR10_MLP_LeNet}) and for the LeNet (see the red line in Figure~\ref{fig:CIFAR10_MLP_LeNet}), considering a global analysis of the parameters. Note, only 30 bit-flips are sufficient to completely crush the accuracy of the two considered networks.

\textbf{Analysis for the CIFAR10 Dataset}: Similar experiments have been performed also for the CIFAR10 dataset~\cite{cifar10}, which is composed of 60,000 training and 10,000 test RGB 32x32 images. The CNN used in our experiments, whose structure is reported in Table~\ref{tab:Cifar10Net}, reaches 79\% of accuracy after 50 epochs of training.

\begin{table}[h!]
\vspace*{-5pt}
	\caption{CNN structure providing 79\% accuracy on CIFAR10.}	
	\vspace*{-3pt}
	\centering
	\begin{tabularx}{9cm}{lcccc}
		\specialrule{.2em}{.1em}{.1em}
		\textbf{Layer} & \textbf{Output shape} & \textbf{Output maps} & \textbf{Kernel size} & \textbf{Strides}\\
		\specialrule{.2em}{.1em}{.1em}
		Input  & (32, 32, 3) & - & - & -\\
		Conv2D  & (32, 32, 32) & 32 & (3,3) & (1,1)\\
		Conv2D  & (30, 30, 32) & 32 & (3,3) & (1,1)\\
		MaxPool2D  & (15, 15, 32) & - & - & (2,2)\\
		Dropout 0.25 & (15, 15, 32) & - & - & - \\
		Conv2D  & (15, 15, 64) & 64 & (3,3) & (1,1)\\
		Conv2D  & (13, 13, 64) & 64 & (3,3) & (1,1)\\
		MaxPool2D  & (6, 6, 64) & - & - & (2,2)\\
		Dropout 0.25 & (6, 6, 64) & - & - & - \\
		Dense  & 512 & - & - & -\\
		Dropout 0.25 & 512 & - & - & - \\
		Dense  & 10 & - & - & -\\
		\specialrule{.2em}{.1em}{.1em}
	\end{tabularx}
	\label{tab:Cifar10Net}
	\vspace*{-5pt}
\end{table}
The \textit{gradient search algorithm} is applied on all the parameters of the network, and similar results w.r.t. the previous cases are obtained. However, as shown by the orange line in Figure~\ref{fig:CIFAR10_MLP_LeNet}, the accuracy drop is far more emphatic. In fact, the accuracy reaches a plateau around 10\% for just 4 bit-flips, which is a more critical result than the one obtained with the LeNet and the MLP working on the MNIST dataset.

\section{NeuroAttack Methodology}
\label{sec:methodology}

\subsection{Threat Model}\label{subsec:ThreatModel}

The attack phase is supposed to be within the supply chain where a malicious actor can insert hardware Trojans. In fact, modern integrated circuit design often involves a number of design houses, fabrication houses, third-party IP, and electronic design automation tools that are all supplied by different vendors. Such a horizontal business model makes the security extremely difficult to manage during the supply chain~\cite{Abbassi2018TrojanZero}\cite{HWTRJN_CNN}. Moreover, the attack is in a \textit{grey-box} setting, i.e., the attacker has a complete knowledge of the system architecture and internal parameters but is not aware of the training set and training hyperparameters. 

\subsection{Hardware Trojan Design}
\begin{figure*}[ht!]
\centering
\includegraphics[width=\linewidth]{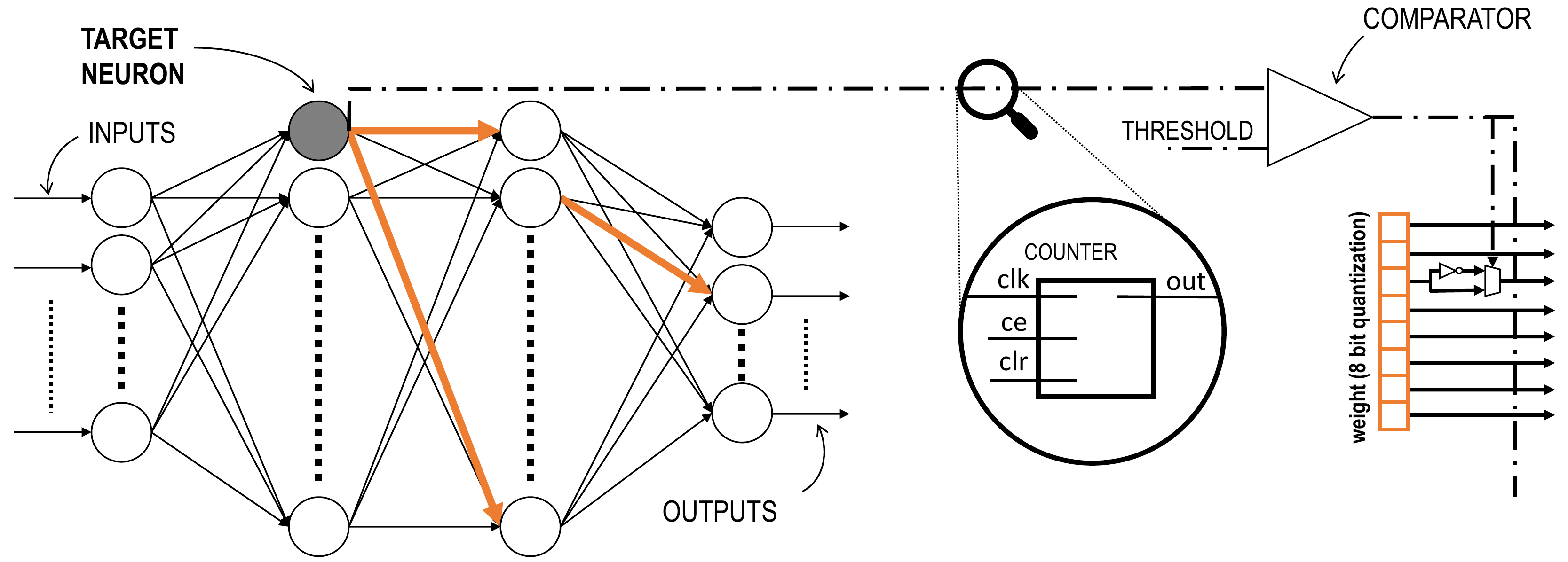}
\caption{Scheme of the Trojan attack for the MLP network with the counter added present only in SNN implementation.}
\label{fig:methodology}
\vspace*{-15pt}
\end{figure*}

The hardware Trojan is designed to perform fault-injection (i.e., bit-flips) in the network parameters to undermine its integrity and degrade its accuracy. The malicious behavior is triggered from the input through a specifically crafted input noise.
The idea is to trigger a fewer number of hardware Trojans hidden in the circuit during the supply chain. Taking advantage of the analysis carried on in Section~\ref{subsec:grad_desc}, hardware stealthy Trojans are inserted at appropriate locations. Each Trojan consists of a 2-way multiplexer with one input which is the original bit, whereas the other input is the complemented bit obtained through an inverter. The multiplexer's selection signal is a signal which is at logic value high only when a trigger is added to the input image. In this way, the network will behave correctly when an untouched input is supplied, providing high accuracy for the original dataset. However, when a trigger is inserted in the input image in form of hidden noise, the fault-injections will be activated, and therefore the accuracy will be degraded significantly. The setting is explained in Figure~\ref{fig:methodology}, in which the thick orange arrows represent the synapses with bit-flip applied, and the grey neuron is the target neuron. To produce the selection signal of the multiplexers, the output of a selected neuron is compared against a threshold through a comparator, chosen according to the results of our experiments. Note, the goal is for the output of the neuron to exceed the threshold when the trigger is added to the dataset, and not when the original dataset is given as an input. The first step of the work is to select a particular neuron to satisfy the desired behavior. To transfer the methodology from the DNN to the SNN domain, a counter that accumulates the number of spikes is needed at the input of the comparator. Moreover, the threshold must be transferred from its analog value to the corresponding value of spike rates. The counter is cleared at the end of the processing of each input.

\subsection{Trigger Pattern Design}
\label{subsec:trigger_pattern}

Since there can be a direct relationship between the analog output value of a neuron and the corresponding spike rate, the knowledge obtained through the analysis of the DNN can be transferred to the SNN implementation. Moreover, a good correlation between analog output value and spike rate is a necessary condition when using the SNN toolbox for DNN-to-SNN conversion. Our goal is to embed the trigger inside one neuron of the network, which we call the \textit{target neuron}. In other words, the goal of our proposed technique is that such a target neuron is activated by a carefully designed mask in the input image.
\subsubsection{Choosing the target layer}
\begin{figure}[b]
\centering
\vspace*{-12pt}
\includegraphics[width=90mm]{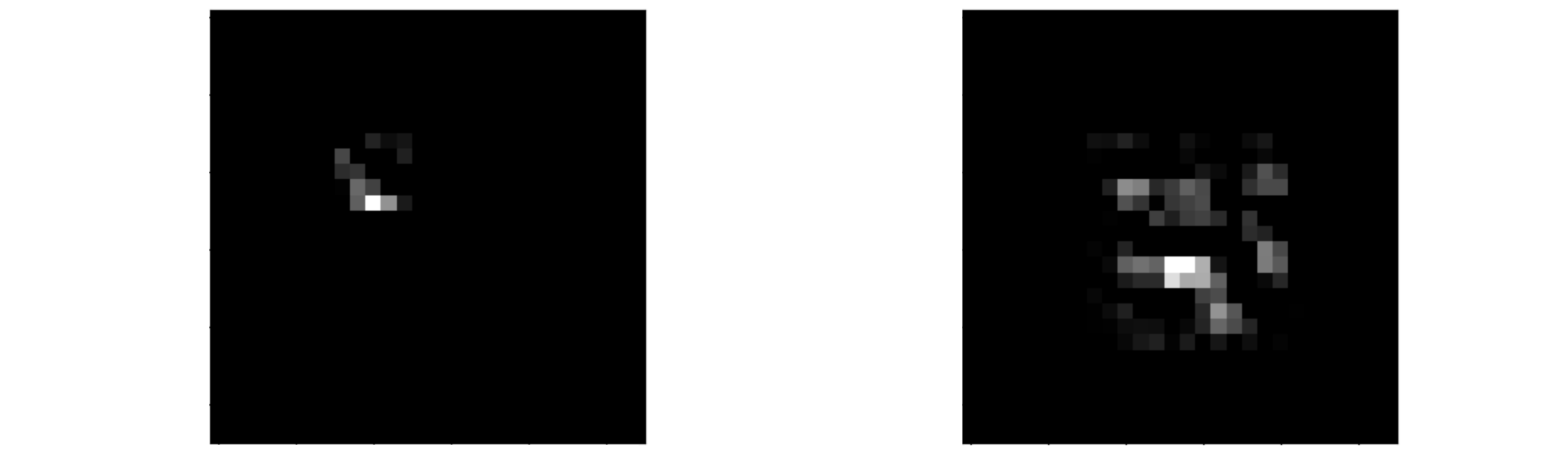}
\caption{Gradient representation of a random neuron from \textbf{(left)} the first and \textbf{(right)} the second convolutional layer of the LeNet.}
\label{fig:merge_grads}
\end{figure}

The selection of the target neuron strongly depends on the target layer. In case of a CNN, the choice of the layer is directly connected to the choice of the size of the trigger mask. This is due to the fact that neurons belonging to deeper convolutional layers are related to a larger area of the input image. For example, by looking at Figure~\ref{fig:merge_grads}, the gradients of a neuron belonging to the first and second convolutional layers are reported. The higher the order of the layer is, the larger the area of the image that will account for the trigger. At the first convolutional layer, the shape, position and value of the gradients are quite clear, and corresponds to the \textit{feature map} of the neurons. For neural networks which have only dense layers (e.g., MLPs) the gradients cover the entire image. In this case, if a smaller trigger is desired, a mask that does not comprehend all the area covered by the gradients can be crafted.

\subsubsection{Choosing the target neuron}

The target neuron is chosen as the one with thehighest value among the sum of absolute values of weights connected to the neurons of the previous layer. This is modeled by the following equation:

\[
argmax_{t}(\sum_{i=1}^N ABS(W_{layer_{i,t}})).
\]

\subsubsection{Choosing the triggering mask}

A \textit{random initial image} is created and the network is inferred with that image, leading to a value $initial_{OUTPUT_{k}}$ at the output of the target neuron. The parameter $target_{OUTPUT_{k}}$ is chosen to be much higher than $initial_{OUTPUT_{k}}$.
A cost function is then defined as follows:

\[
cost = \dfrac{\sum_{i=1}^N \delta_{i}^{2}}{N},
\]
Where $\delta_{i} = target_{OUTPUT_{i}}-initial_{OUTPUT_{i}}$, \textit{i} is the index of each neuron in the target layer. Being \textit{k} the index of the target neuron, we rewrite the expression as:

\[
cost = \dfrac{\delta_{1}^{2} + \delta_{2}^{2} + ... + \delta_{k}^{2} + ... + \delta_{N}^{2}}{N}
\]
For each $\delta_{i}$ it is imposed that $target_{OUTPUT_{i}} = initial_{OUTPUT_{i}}$ except for $\delta_{k}$, where $target_{OUTPUT_{k}} \neq initial_{OUTPUT_{k}}$. The derivative of the cost function is computed with respect to the pixels of the random input image, to understand which part of the input image influences the target neuron. Based on this, a mask \textit{M} is created and a \textit{random initial trigger} is generated by the dot product between the mask and the \textit{random initial image}. The mask can also be chosen differently, but it must have some overlap with the gradient matrix, otherwise the loop that has to be described, will not work.

\subsubsection{Generating the trigger}

The trigger generation algorithm (see Algorithm~\ref{code:trigger_gen}) is inspired by the work of Liu et al. on Trojan attacks~\cite{inproceedings}. In the first rows, some initialization parameters are set. $val_{min}$ and $val_{max}$ are useful to manage the imperceptibility characteristics of the trigger, but should always lay in the range (0,1). The loop proceeds until the cost reaches a particular threshold, or until a maximum number of iterations. The gradients $\Delta$ are first calculated and then limited by a mask that can be suited for the gradients (in that case, line 4 of Algorithm~\ref{code:trigger_gen} can be skipped), or can be decided in another way. Compared to algorithm in~\cite{inproceedings}, line 6 is added to limit the maximum and minimum values for the pixels in the trigger.

\begin{algorithm}
\caption{Trigger generation loop}\label{code:trigger_gen}
\begin{algorithmic}[1]
\State \textbf{INIT}$(val_{min}, val_{max}, lr,  epc, epochs, th, cost)$
\While {$\textit{cost} < \textit{th}$ and $\textit{epc} < \textit{epochs}$}
\State $\Delta = \dfrac{\partial cost}{\partial x}$
\State $\Delta = \Delta \cdot M$
\State $x = x - lr\cdot \Delta$
\State $x = clip(x, val_{min}, val_{max})$
\State $epc = epc + 1$
\EndWhile
\State \textbf{return} x;
\end{algorithmic}
\end{algorithm}

At the end of the loop, a new trigger is generated with pixels' values optimized to provoke the saturation of the target neuron. If the parameter $target_{value_{k}}$ is set too high, in general, the target neuron will not reach that value but a lower value, which we call $final_{value_{k}}$. A \textit{threshold} is chosen, such as that if the neuron's output value exceeds it, the output of the comparator is set to high and the multiplexers are switched. Then, for each targeted weight, the selected bit is complemented. The \textit{threshold} is calculated through the following formula:
\[
threshold = final_{value_{k}} - \xi,
\]
where $\xi$ is a parameter, which can be chosen according to the parallelism of the network and the method of the attack. 

\subsubsection{Trigger application}\label{trigger_app_methods}

The trigger can be applied on the image in mainly two ways: (1) as a stamp in the image, or (2) as a noise in the image. In the first case, the values of the pixel in the trigger area are exactly the optimal ones as generated by the loop described in the lines 2-7 of Algorithm~\ref{code:trigger_gen}. However, this solution could be less imperceptible, and in that case a careful choice of the layer and/or a careful choice of the trigger mask parameters (position, dimension, $max_{val}$) should be taken into consideration. The second case could be of a more general interest and it produced good results, due to a better imperceptibility, as it will be shown in the following Section~\ref{sec:results}. Moreover, supposing to have some general knowledge about the pixel intensity distribution on the image dataset targeted by the network, the choice of the trigger parameters can rely also on this information.

\section{Results and Discussion}
\label{sec:results}

\subsection{Experimental Setup}

Both the original and the modified dataset are used for inference, and the amount of times for which both dataset make the target neuron exceed the threshold is recorded. There is the possibility that some images from the original dataset produce the saturation of the neuron, causing an unwanted activation of the Trojans for an $exceed_{ORIGINAL}$ amount of times. However, for a stealthy attack purpose, a carefully crafted trigger should lead to a situation in which this value is kept to almost zero. Therefore, the accuracy is not noticeably reduced when the input trigger is not present, i.e., the presence of hardware Trojans is stealthy. We call $dim_{DATASET}$ the number of images in the dataset, $exceed_{ORIGINAL}$ the number of images from the original dataset for which the threshold for the target neuron is exceeded, and $exceed_{MODIFIED}$ the number of images from the modified dataset for which the threshold for the target neuron is exceeded. Hence, the attack aims at being both effective and stealthy, and thereby to simultaneously satisfy the following conditions:

\begin{enumerate}
\item $exceed_{ORIGINAL} << exceed_{MODIFIED}$
\item $exceed_{ORIGINAL} << dim_{DATASET}$
\item $exceed_{MODIFIED} \simeq dim_{DATASET}$
\end{enumerate}

In the following, the results obtained using the MNIST and the CIFAR10 datasets are discussed.

\subsubsection{Results on the MNIST dataset}

\begin{figure}[ht!]
\centering
\vspace*{-5pt}
\includegraphics[width=90mm]{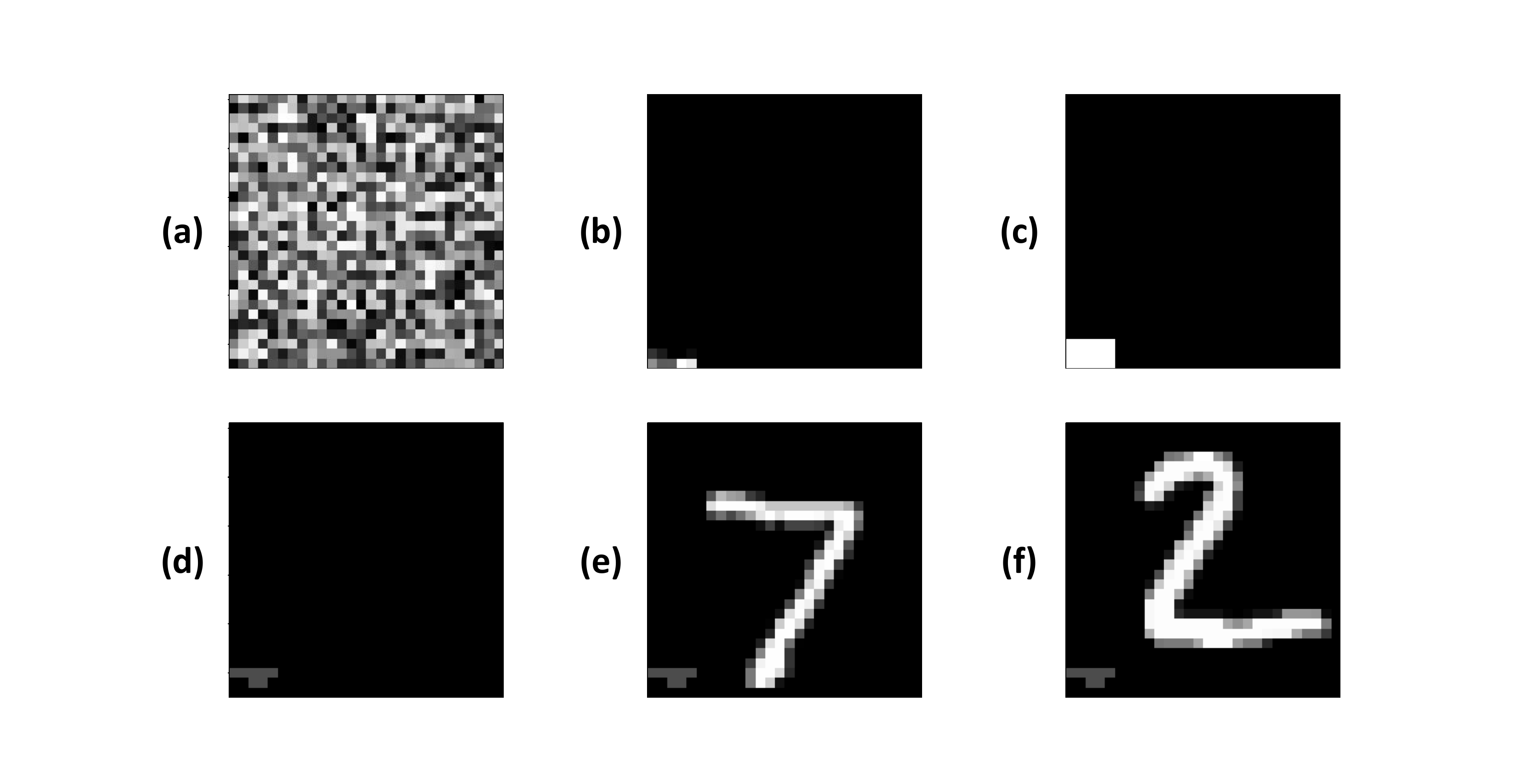}
\vspace*{-20pt}
\caption{From top-left to bottom-right: (a) initial input trigger, (b) gradients of the selected neuron, (c) mask created through gradients, (d) final trigger after loop, (e) and (f) two images with applied trigger.}
\label{fig:6subplt_layer1}
\vspace*{-5pt}
\end{figure}

Targeting the \textbf{first convolutional layer} of the LeNet-5 with parameters listed in the first row of Table~\ref{tab:trigger_parameters}, the trigger shown in Figures~\ref{fig:6subplt_layer1} (d) is produced. 

In Figure~\ref{fig:6subplt_layer1} (a), (b) and (c), the \textit{random initial image}, the initial gradients and the mask \textit{M} are shown respectively. The mask is crafted to follow the shape of the gradients. The images from both the original and the modified test set (two examples from this last image set are shown in Figures~\ref{fig:6subplt_layer1} (e) and (f)) are inferred and the results, as reported in Table~\ref{tab:trigger_parameters}: $exceed_{ORIGINAL} = 0$ and $exceed_{MODIFIED} = 10000$.

\begin{figure}[ht!]
\centering
\vspace*{-12pt}
\includegraphics[width=90mm]{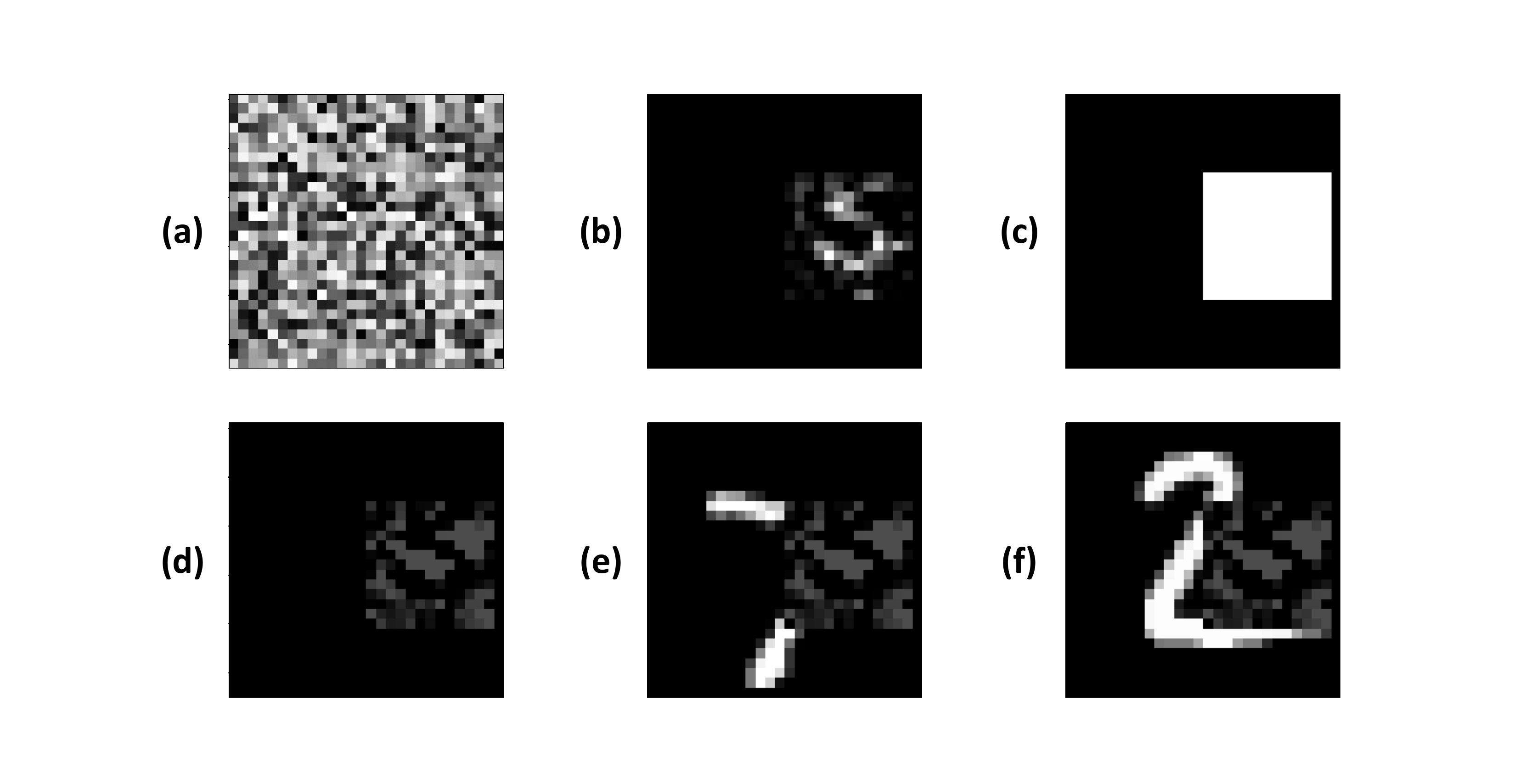}
\vspace*{-20pt}
\caption{From top-left to bottom-right: (a) initial input trigger, (b) gradients of the selected neuron, (c) mask created through gradients, (d) final trigger after loop, (e) and (f) two images with applied trigger.}
\label{fig:MNIST_LeNet_2nd_layer}
\vspace*{-5pt}
\end{figure}
Targeting the \textbf{second convolutional layer}, the produced results are significantly different. In fact, the trigger is far more perceptible and superimposed with a significant part of the images, as can be seen in Figure~\ref{fig:MNIST_LeNet_2nd_layer}. In this case, with the same experimental settings as explained earlier, the obtained statistics about the threshold exceeding are: $exceed_{ORIGINAL} = 5$ and $exceed_{MODIFIED} = 7585$, as also reported in Table~\ref{tab:trigger_parameters}.
This demonstrates that targeting a neuron belonging to the second convolution layer leads to a relatively worse result. In fact, it can be pointed out that the gradients are, on average, higher than the gradients corresponding to a target neuron belonging to the first convolution layer. We define the correlation between the target neuron and the masked part of the image \textit{S} as follows: 
\[
S = \dfrac{\sum_{i,j}^N \gamma_{i,j}}{M^{2}},
\]

Where $\gamma_{i,j}$ is the gradient corresponding to the pixel with indexes \textit{i,j} in the trigger mask, and \textit{M} is the size of the side trigger, in case of a square trigger. It can be seen that in the first convolution layer $S=2.21 \cdot 10^{-5}$, whereas in the second convolution layer $S=1.4 \cdot 10^{-6}$. This clearly shows that, for a neuron in the $2^{nd}$ layer, the variation with the input pixel is much lower.
If we call $\rho$ the value
\[
\rho = exceed_{MODIFIED} - exceed_{ORIGINAL},
\]
we can see that it is getting lower when choosing target neurons belonging to deeper layers.

\begin{figure}[ht!]
\centering
\vspace*{-10pt}
\includegraphics[width=90mm]{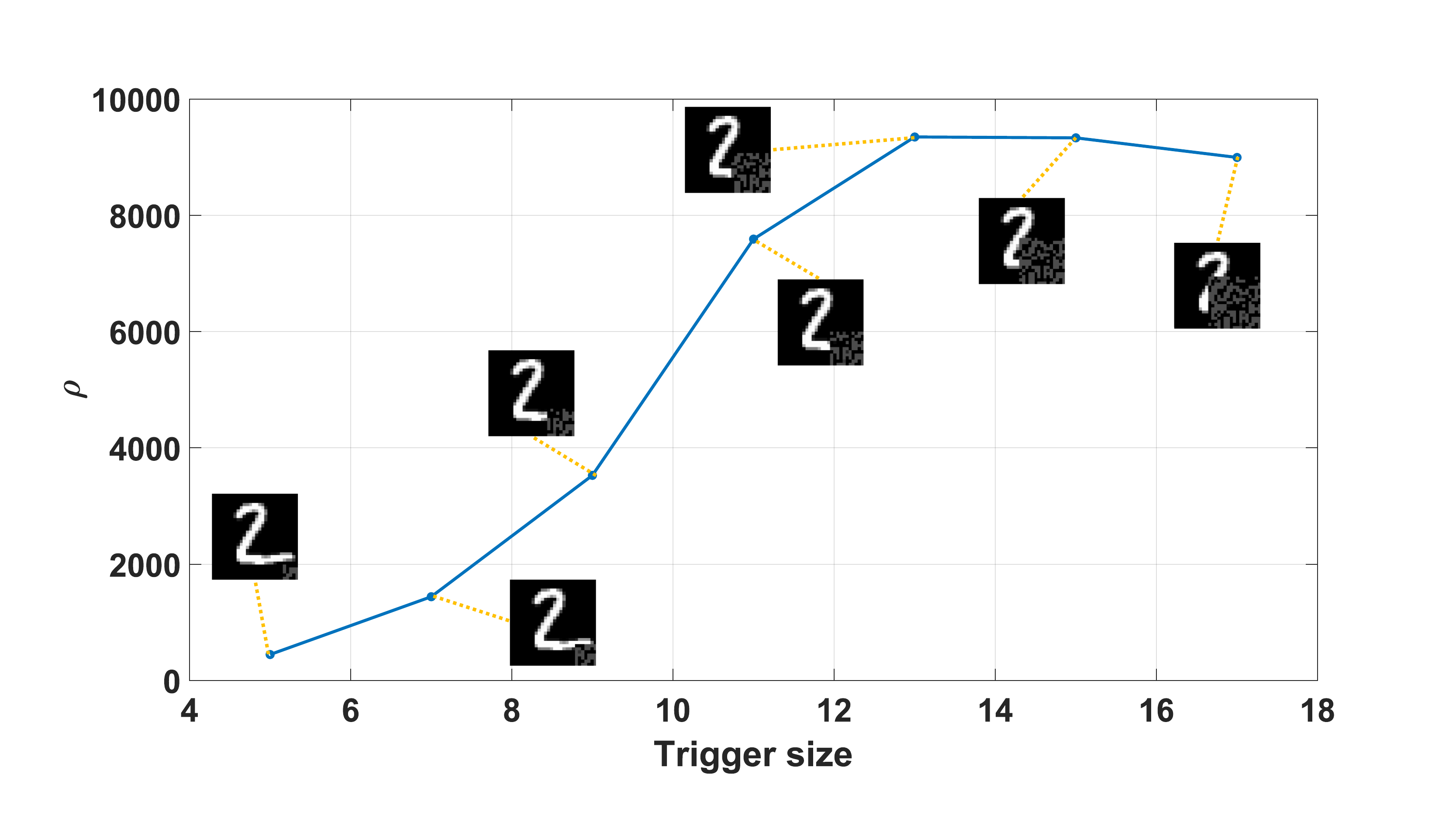}
\vspace*{-15pt}
\caption{Plot of $\rho$ with respect to the trigger size.}
\label{fig:rho}
\end{figure}

Taking into consideration the MLP, a square mask is created and put in the bottom-right corner. Its side is varied between 5 and 17 pixels, with steps of 2 pixels. Since, at the beginning, the area of the trigger is too small, there are not enough pixels to optimize the saturation of the target neuron. The difference between $initial_{value_{k}}$ and $final_{value_{k}}$ results in a small value. Moreover, a huge number of images from the original dataset make the target neuron exceed the threshold, leading to a small value of $\rho$. A larger area of the trigger, on one hand, increases $\rho$ as can be seen in Figure~\ref{fig:rho} and, on the other hand, leads to a less stealthy trigger.

\begin{figure}[ht!]
\centering
\vspace*{-14pt}
\includegraphics[width=90mm]{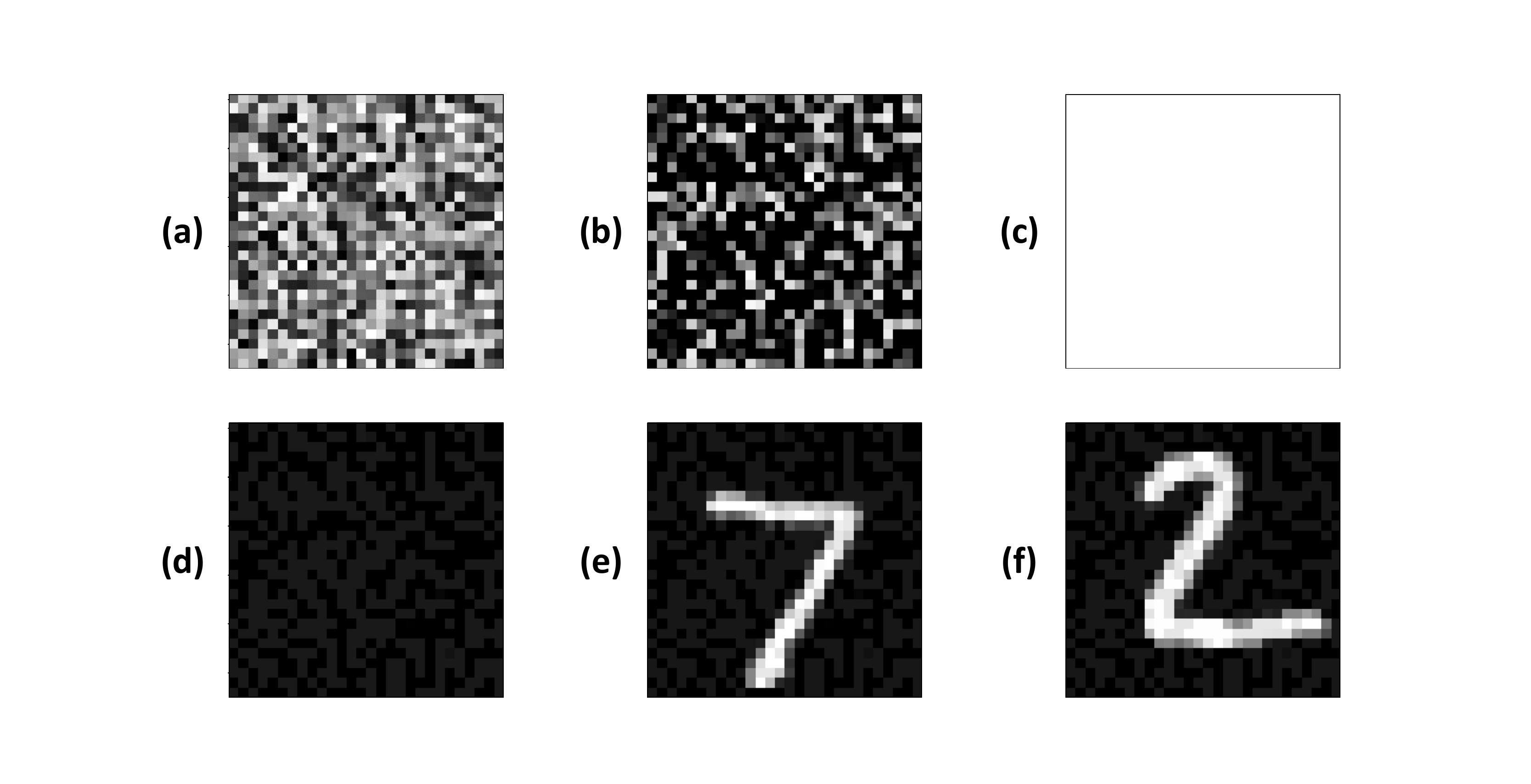}
\vspace*{-15pt}
\caption{From top-left to bottom-right (a) initial input trigger, (b) gradients of the selected neuron, (c) mask created through gradients, (d) final trigger after loop, (e) and (f) two images with applied trigger.}
\label{fig:MLP_trigger}
\end{figure}

In the case of the MLP network, an interesting result is obtained with a lower value of $max_{val}=0.1$. Even though we are targeting the first layer, the gradients are covering the complete image (Figure~\ref{fig:MLP_trigger} (b)), since it is a fully-connected layer. Hence, we create a mask suited for the gradient, which spans across the whole image, as shown in Figure~\ref{fig:MLP_trigger} (c). In this case, the second method described in Section~\ref{trigger_app_methods} is used to apply the trigger. Due to the low value of $val_{max}$, the trigger results to be imperceptible, as shown in Figures~\ref{fig:MLP_trigger} (e) and (f). We obtained a very high $\rho$, shown in Table~\ref{tab:trigger_parameters}, and high imperceptibility, at the expense of a harder applicability.

\subsubsection{Results on the CIFAR10 dataset}
\begin{figure}[ht!]
\centering
\vspace*{-14pt}
\includegraphics[width=90mm]{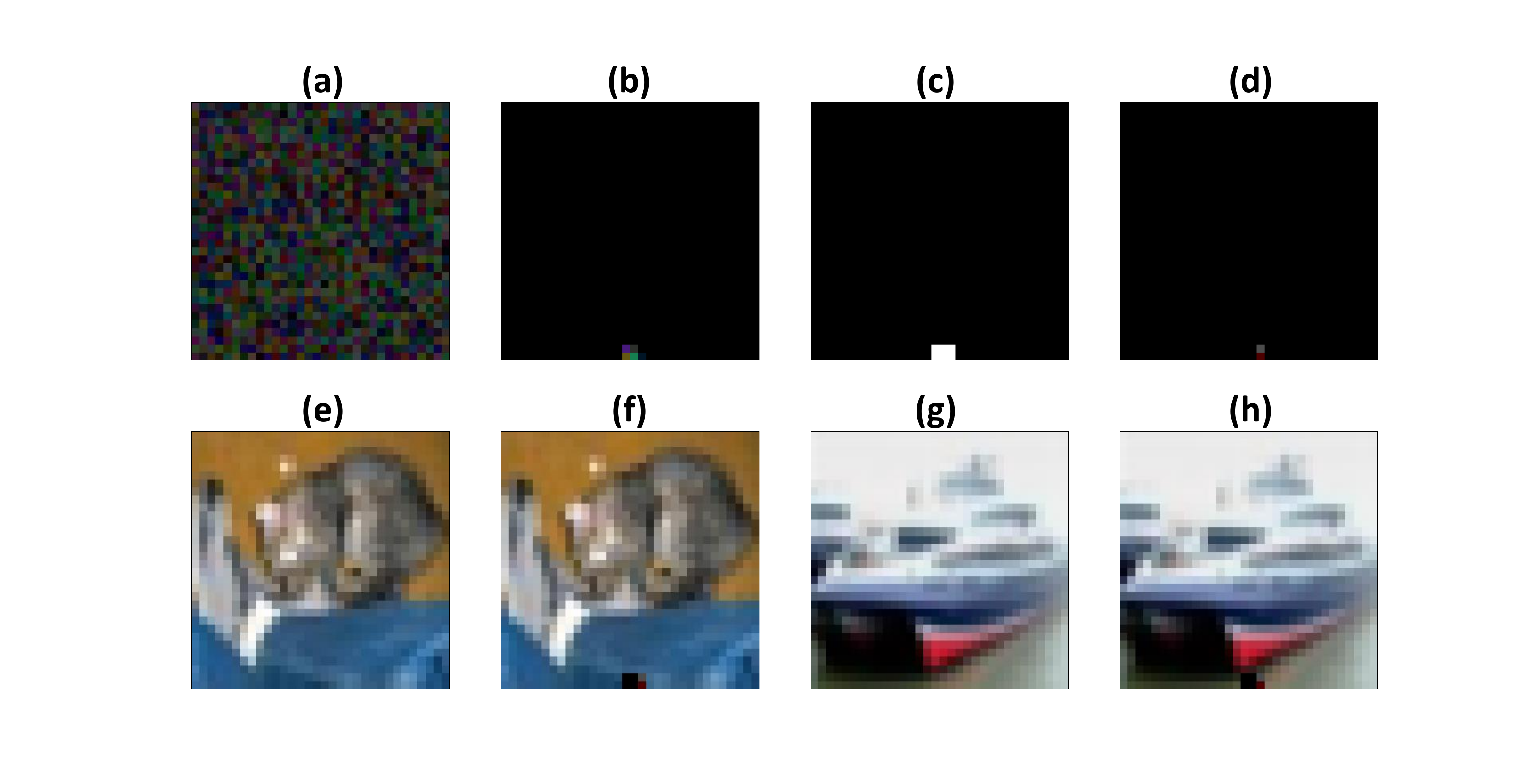}
\vspace*{-15pt}
\caption{From top-left to bottom-right: (a) initial input trigger, (b) gradients of the selected neuron, (c) mask created through gradients, (d) final trigger after loop, (e) first image from the dataset (f) first image with trigger applied (g) second image from the dataset (h) second image with trigger applied.}
\label{fig:CIFAR10merge}
\vspace*{-6pt}
\end{figure}

\begin{table*}[h!]
	\caption{Structure of the networks, parameters and results for our experiments.}	
	\centering
	\begin{tabularx}{18.2 cm}{cc|ccccc|cc}
		\specialrule{.2em}{.1em}{.1em}
		\textbf{Net} & \textbf{Layer} & \textbf{$val_{max}$} & \textbf{$\xi$} & \textbf{$target_{OUTPUT_{k}}$} & \textbf{$initial_{VAL_{k}}$} & \textbf{$final_{VAL_{k}}$} & \textbf{$exceed_{ORIGINAL}$} & \textbf{$exceed_{MODIFIED}$}\\
		\specialrule{.2em}{.1em}{.1em}
		MNIST LeNet & 1st Conv2D  & 0.3 & 0.1 & 100 & 0.04 & 0.21 & 0 & \textbf{10000}\\
		\hline
		MNIST LeNet & 2nd Conv2D  & 0.3 & 0.1 & 100 & 0.08 & 1.56 & 5 & 7585\\
		\hline
		MNIST MLP & 1st Dense  & 0.1 & 0.1 & 100 & 0.05 & 1.21 & 15 & 9904\\
		\hline
		CIFAR10 CNN & 1st Conv2D  & 0.3 & 0.1 & 100 & 0.02 & 0.23 & 4 & \textbf{10000}\\
		\specialrule{.2em}{.1em}{.1em}
	\end{tabularx}
	\label{tab:trigger_parameters}
	\vspace*{-10pt}
\end{table*}
In this case, targeting the first layer, with parameters set as shown in Table~\ref{tab:trigger_parameters}, the trigger shown in Figure~\ref{fig:CIFAR10merge} (d) is produced. The superposition of the trigger on the original images (two examples) is shown in Figures~\ref{fig:CIFAR10merge} (f) and (h)).

\subsection{Hardware Overhead}\label{subsec:HWOverhead}
Given the amount M of bit-flips applied, the hardware overhead is constituted as the following.
\begin{enumerate}
\item M inverters, constituted by 2 transistors each.
\item M 2-way multiplexer, constituted by 16 transistors each in a 4 NANDs implementation.
\item \textit{In the case of a DNN}, a digital comparator, whose complexity depends on the parallelism of the neuron's output result, which is connected to the target neuron's output.
\item \textit{In the case of an SNN}, a counter, to count the spikes, plus a compartor which is set when the counter reaches a particular value.
\end{enumerate}
The overhead of multiplexers and inverters can be estimated as $(2+16) \times M $. From the experiments reported in Section~\ref{subsec:grad_desc}, it is clear that an amount of about just 30 bit-flips is enough to completely crash the performances of the DNN for the two networks operating on MNIST dataset, or 4 bit-flips in the case of the CNN operating on the CIFAR10 dataset. The hardware overhead of inverters and multiplexers, calculated in terms of transistors, is about $(2+16) \times 30 = 540$ in the first case, whereas it is just $(2+16) \times 4 = 72$ in the second case. In the case of a SNN, a counter is added, whose module should be at least as much as the maximum spiking rate a neuron can have. The amount of transistors needed for a module N counter are given by  $\#_{transistors} = (N-2) \times 6 + (N \times 4) \times 4$, where the first addend gives the contribution of the AND gates, whereas the second gives the contribution of the T-type flip-flops.

\section{Conclusion}

In this paper, we propose \textit{NeuroAttack}, a cross-layer attack against DNNs and SNNs, that exploits a circuit-level vulnerability to threaten security. In particular, we demonstrated that NeuroAttack can drastically degrade the accuracy of a DNN or an SNN by applying a few number of bit-flips on its parameters, through a hardware Trojan triggered externally by an adversarial input noise. The security issue is made more severe by the stealthiness of the attack, since it is only effective when triggered by the external adversarial noise, and practically imperceptible elsewhere. Due to the linear relationship between DNN activations and SNN spike rates, the obtained results are transferred to SNN models to corroborate the fact that the demonstrated attack presents a clear threat to both SNNs and DNNs. 

\section*{Acknowledgments}

\begin{small}
\noindent
This work has been partially supported by the Doctoral College Resilient Embedded Systems which is run jointly by TU Wien's Faculty of Informatics and FH-Technikum Wien.
\end{small}

\begin{refsize}
\bibliographystyle{abbrvnat}
\bibliography{main.bib}

\begin{thebibliography}{44}
\providecommand{\natexlab}[1]{#1}
\providecommand{\url}[1]{\texttt{#1}}
\expandafter\ifx\csname urlstyle\endcsname\relax
  \providecommand{\doi}[1]{doi: #1}\else
  \providecommand{\doi}{doi: \begingroup \urlstyle{rm}\Url}\fi

\bibitem[Abbassi et~al.(2018)]{Abbassi2018TrojanZero}
I.~H. Abbassi et~al.
\newblock Trojanzero: Switching activity-aware design of undetectable hardware
  trojans with zero power and area footprint.
\newblock \emph{DATE}, 2018.

\bibitem[Beeman(2013)]{Hodgkin-Huxley}
D.~Beeman.
\newblock Hodgkin-huxley model.
\newblock In \emph{Encyclopedia of Computational Neuroscience}, 2013.

\bibitem[Bohte et~al.(2002)Bohte, Kok, and Poutré]{BOHTE200217}
S.~M. Bohte, J.~N. Kok, and H.~L. Poutré.
\newblock Error-backpropagation in temporally encoded networks of spiking
  neurons.
\newblock \emph{Neurocomputing}, 2002.

\bibitem[Bouvier et~al.(2019)]{Bouvier}
M.~Bouvier et~al.
\newblock Spiking neural networks hardware implementations and challenges: A
  survey.
\newblock 2019.

\bibitem[Cardaliaguet and Euvrard(1992)]{STDP}
P.~Cardaliaguet and G.~Euvrard.
\newblock Approximation of a function and its derivative with a neural network.
\newblock \emph{Neural Networks}, 1992.

\bibitem[Clements and Lao(2018)]{HWTRJN_CNN}
J.~Clements and Y.~Lao.
\newblock Hardware trojan attacks on neural networks.
\newblock \emph{CoRR}, abs/1806.05768, 2018.

\bibitem[{Davies} et~al.(2018)]{Davies2018Loihi}
M.~{Davies} et~al.
\newblock Loihi: A neuromorphic manycore processor with on-chip learning.
\newblock \emph{IEEE Micro}, 2018.

\bibitem[{Gibson} et~al.(1989){Gibson}, {Siu}, and {Cowen}]{266645}
G.~J. {Gibson}, S.~{Siu}, and C.~F.~N. {Cowen}.
\newblock Multilayer perceptron structures applied to adaptive equalisers for
  data communications.
\newblock In \emph{ICASSP}, 1989.

\bibitem[Goodfellow et~al.(2014)Goodfellow, Shlens, and Szegedy]{Goodfellow}
I.~J. Goodfellow, J.~Shlens, and C.~Szegedy.
\newblock Explaining and harnessing adversarial examples, 2014.

\bibitem[Han et~al.(2016)Han, Mao, and Dally]{Han2016DeepCompression}
S.~Han, H.~Mao, and W.~J. Dally.
\newblock Deep compression: Compressing deep neural network with pruning,
  trained quantization and huffman coding.
\newblock In \emph{ICLR}, 2016.

\bibitem[Hanif and Shafique(2019)]{Hanif2019SalvageDNN}
M.~A. Hanif and M.~Shafique.
\newblock Salvagednn: salvaging deep neural network accelerators with permanent
  faults through saliency-driven fault-aware mapping.
\newblock \emph{Philosophical Transactions of the Royal Society A}, 2019.

\bibitem[{Hanif} et~al.(2018){Hanif}, {Hafiz}, and {Shafique}]{8342139}
M.~A. {Hanif}, R.~{Hafiz}, and M.~{Shafique}.
\newblock Error resilience analysis for systematically employing approximate
  computing in convolutional neural networks.
\newblock In \emph{DATE}, 2018.

\bibitem[Hanif et~al.(2018{\natexlab{a}})]{Hanif2018RobustML}
M.~A. Hanif et~al.
\newblock Robust machine learning systems: Reliability and security for deep
  neural networks.
\newblock \emph{IOLTS}, 2018{\natexlab{a}}.

\bibitem[Hanif et~al.(2018{\natexlab{b}})]{Hanif2018X-DNNs}
M.~A. Hanif et~al.
\newblock X-dnns: Systematic cross-layer approximations for energy-efficient
  deep neural networks.
\newblock \emph{J. Low Power Electronics}, 2018{\natexlab{b}}.

\bibitem[Hazan et~al.(2018)]{Hazan_2018}
H.~Hazan et~al.
\newblock Bindsnet: A machine learning-oriented spiking neural networks library
  in python.
\newblock \emph{Frontiers in Neuroinformatics}, 2018.

\bibitem[Hoang et~al.(2020)Hoang, Hanif, and Shafique]{Hoang2020FTClipAct}
L.-H. Hoang, M.~A. Hanif, and M.~Shafique.
\newblock Ft-clipact: Resilience analysis of deep neural networks and improving
  their fault tolerance using clipped activation.
\newblock In \emph{DATE}, 2020.

\bibitem[{Izhikevich}(2003)]{1257420}
E.~M. {Izhikevich}.
\newblock Simple model of spiking neurons.
\newblock \emph{IEEE Transactions on Neural Networks}, 2003.

\bibitem[{Kim} et~al.(2014)]{rowhammer}
Y.~{Kim} et~al.
\newblock Flipping bits in memory without accessing them: An experimental study
  of dram disturbance errors.
\newblock In \emph{ISCA}, 2014.

\bibitem[Krizhevsky(2012)]{cifar10}
A.~Krizhevsky.
\newblock Learning multiple layers of features from tiny images.
\newblock \emph{University of Toronto}, 2012.

\bibitem[LeCun et~al.(1998)]{LeCun1998LeNet}
Y.~LeCun et~al.
\newblock Gradient-based learning applied to document recognition.
\newblock 1998.

\bibitem[Lee et~al.(2019)Lee, Sarwar, and
  Roy]{DBLP:journals/corr/abs-1903-06379}
C.~Lee, S.~S. Sarwar, and K.~Roy.
\newblock Enabling spike-based backpropagation in state-of-the-art deep neural
  network architectures.
\newblock \emph{CoRR}, abs/1903.06379, 2019.

\bibitem[Lee et~al.(2018)]{Lee2018CNNSTDP}
C.~Lee et~al.
\newblock Training deep spiking convolutional neural networks with stdp-based
  unsupervised pre-training followed by supervised fine-tuning.
\newblock \emph{Frontiers in Neuroscience}, 2018.

\bibitem[Lee et~al.(2016)Lee, Delbr{\"{u}}ck, and
  Pfeiffer]{DBLP:journals/corr/LeeDP16}
J.~Lee, T.~Delbr{\"{u}}ck, and M.~Pfeiffer.
\newblock Training deep spiking neural networks using backpropagation.
\newblock \emph{CoRR}, abs/1608.08782, 2016.

\bibitem[{Liu} et~al.(2017)]{FICNN}
Y.~{Liu} et~al.
\newblock Fault injection attack on deep neural network.
\newblock In \emph{ICCAD}, 2017.

\bibitem[Liu et~al.(2018)]{inproceedings}
Y.~Liu et~al.
\newblock Trojaning attack on neural networks.
\newblock 2018.

\bibitem[{Marchisio} et~al.(2018){Marchisio}, {Hanif}, {Martina}, and
  {Shafique}]{Marchisio2018PruNet}
A.~{Marchisio}, M.~A. {Hanif}, M.~{Martina}, and M.~{Shafique}.
\newblock Prunet: Class-blind pruning method for deep neural networks.
\newblock In \emph{IJCNN}, 2018.

\bibitem[Marchisio et~al.(2020{\natexlab{a}})Marchisio, Mrazek, Hanif, and
  Shafique]{Marchisio2020ReD-CaNe}
A.~Marchisio, V.~Mrazek, M.~A. Hanif, and M.~Shafique.
\newblock Red-cane: A systematic methodology for resilience analysis and design
  of capsule networks under approximations.
\newblock In \emph{DATE}, 2020{\natexlab{a}}.

\bibitem[{Marchisio} et~al.(2019)]{Marchisio2019DL4EC}
A.~{Marchisio} et~al.
\newblock Deep learning for edge computing: Current trends, cross-layer
  optimizations, and open research challenges.
\newblock In \emph{ISVLSI}, 2019.

\bibitem[Marchisio et~al.(2020{\natexlab{b}})]{Marchisio2019SNNAttack}
A.~Marchisio et~al.
\newblock Is spiking secure? a comparative study on the security
  vulnerabilities of spiking and deep neural networks.
\newblock \emph{IJCNN}, 2020{\natexlab{b}}.

\bibitem[Marchisio et~al.(2020{\natexlab{c}})]{Marchisio2020Q-CapsNets}
A.~Marchisio et~al.
\newblock Q-capsnets: {A} specialized framework for quantizing capsule
  networks.
\newblock \emph{DAC}, 2020{\natexlab{c}}.

\bibitem[Merolla et~al.(2014)]{Merolla2014TrueNorth}
P.~A. Merolla et~al.
\newblock A million spiking-neuron integrated circuit with a scalable
  communication network and interface.
\newblock \emph{Science}, 2014.

\bibitem[{Neftci} et~al.(2019){Neftci}, {Mostafa}, and {Zenke}]{SLAYER}
E.~O. {Neftci}, H.~{Mostafa}, and F.~{Zenke}.
\newblock Surrogate gradient learning in spiking neural networks: Bringing the
  power of gradient-based optimization to spiking neural networks.
\newblock \emph{Signal Processing Magazine}, 2019.

\bibitem[{Neggaz} et~al.(2018){Neggaz}, {Alouani}, {Lorenzo}, and
  {Niar}]{iccd18}
M.~A. {Neggaz}, I.~{Alouani}, P.~R. {Lorenzo}, and S.~{Niar}.
\newblock A reliability study on cnns for critical embedded systems.
\newblock In \emph{ICCD}, 2018.

\bibitem[{Neggaz} et~al.(2019){Neggaz}, {Alouani}, {Niar}, and {Kurdahi}]{dant}
M.~A. {Neggaz}, I.~{Alouani}, S.~{Niar}, and F.~{Kurdahi}.
\newblock Are cnns reliable enough for critical applications? an exploratory
  study.
\newblock \emph{IEEE Design Test}, 2019.

\bibitem[Rakin et~al.(2019)Rakin, He, and
  Fan]{DBLP:journals/corr/abs-1903-12269}
A.~S. Rakin, Z.~He, and D.~Fan.
\newblock Bit-flip attack: Crushing neural network withprogressive bit search.
\newblock \emph{CoRR}, abs/1903.12269, 2019.

\bibitem[{Reagen} et~al.(2016)]{7551399}
B.~{Reagen} et~al.
\newblock Minerva: Enabling low-power, highly-accurate deep neural network
  accelerators.
\newblock In \emph{ISCA}, 2016.

\bibitem[{Reagen} et~al.(2018)]{8465834}
B.~{Reagen} et~al.
\newblock Ares: A framework for quantifying the resilience of deep neural
  networks.
\newblock In \emph{DAC}, 2018.

\bibitem[{Rodriguez} et~al.(2019)]{laserfaultinjection}
J.~{Rodriguez} et~al.
\newblock Llfi: Lateral laser fault injection attack.
\newblock In \emph{FDT)}, 2019.

\bibitem[Rueckauer et~al.()]{10.3389/fnins.2017.00682}
B.~Rueckauer et~al.
\newblock Conversion of continuous-valued deep networks to efficient
  event-driven networks for image classification.
\newblock \emph{Frontiers in Neuroscience}.

\bibitem[{Shafique} et~al.(2020)]{Shafique2020RobustML}
M.~{Shafique} et~al.
\newblock Robust machine learning systems: Challenges,current trends,
  perspectives, and the road ahead.
\newblock \emph{IEEE Design Test}, 2020.

\bibitem[Vaila et~al.(2019)Vaila, Chiasson, and
  Saxena]{DBLP:journals/corr/abs-1903-12272}
R.~Vaila, J.~Chiasson, and V.~Saxena.
\newblock Deep convolutional spiking neural networks for image classification.
\newblock \emph{CoRR}, abs/1903.12272, 2019.

\bibitem[Wang et~al.(2014)Wang, Guo, and Adjouadi]{LIF_neuron_model}
Z.~Wang, L.~Guo, and M.~Adjouadi.
\newblock A generalized leaky integrate-and-fire neuron model with fast
  implementation method.
\newblock \emph{International journal of neural systems}, 2014.

\bibitem[Yuan et~al.(2017)]{pbform}
X.~Yuan et~al.
\newblock Adversarial examples: Attacks and defenses for deep learning.
\newblock \emph{CoRR}, abs/1712.07107, 2017.

\bibitem[Zhang et~al.(2019)]{Zhang2019RobustML}
J.~J. Zhang et~al.
\newblock Building robust machine learning systems: Current progress, research
  challenges, and opportunities.
\newblock \emph{DAC}, 2019.

\end{thebibliography}
\end{refsize}

\end{document}